\documentclass{aip-cp}

\usepackage[numbers]{natbib}
\usepackage{rotating}
\usepackage{graphicx}
\usepackage{wasysym}

\begin{document}

\title{Combined Magnetohydrodynamic- Monte Carlo Simulations of Proton Acceleration in Colliding Wind Binaries}

\author[aff1]{Emanuele Grimaldo\corref{cor1}}
\author[aff1,aff2]{Anita Reimer}
\author[aff2]{Ralf Kissmann}

\affil[aff1]{Institute for Theoretical Physics, University of Innsbruck, Technikerstr. 21A, 6020 Austria.}
\affil[aff2]{Institute for Astro- and Particle Physics, University of Innsbruck, Technikerstr. 25, 6020 Austria.}
\corresp[cor1]{Corresponding author: emanuele.grimaldo@uibk.ac.at}

\maketitle

\begin{abstract}
The interaction between the strong winds of the stars in colliding-wind binary (CWB) systems produces two shock fronts, delimiting the wind collision region (WCR). There, particles are expected to be accelerated mainly via diffusive shock acceleration, and to produce $\gamma$-rays, in processes involving relativistic electrons and/or protons.\\
We investigate the injection and the acceleration of protons in typical CWB systems by means of Monte Carlo simulations, with a test-particle approach. We use magnetohydrodynamic simulations to determine the background conditions in the wind collision region. This allows us to consider particle acceleration at both shocks, on either side of the WCR, with a self-consistently determined large-scale magnetic field, which has an impact on the shape of the WCR, and the topology of which plays an important role in particle acceleration at collisionless shocks. Such studies may contribute to improve $\gamma$-ray flux predictions for CWB systems.
\end{abstract}

\section{INTRODUCTION}
Collisionless shock fronts are known to accelerate particles, primarily \textit{via} first-order Fermi acceleration, resulting in a power law energy spectrum of the non-thermal population (e.g. \cite{jones1991rev, drury1983} and references therein).\\
Colliding-wind binaries (CWBs) - binary systems of massive, hot stars with strong stellar winds - are expected to be sites where particle acceleration occurs. Shocks form at the wind collision region (WCR), thus creating a suitable environment for accelerating particles by means of diffusive shock acceleration (DSA). Besides the detected radio synchrotron radiation from many such systems \cite{williams1997, chapman1999, dougherty2005}, processes such as inverse Compton (IC) scattering, relativistic Bremsstrahlung, and decay of neutral pions produced in hadronic interactions are expected to provide sufficiently high fluxes of GeV and TeV $\gamma$-rays to allow their detection by instruments such as \textit{Fermi}-LAT, HESS, MAGIC and VERITAS (e.g. \cite{eichler1993,reimer2006,benaglia2003}). In contrast to such predictions, $\gamma$-ray emission from CWBs has not been observed up to now, with the exception of $\eta$ Carinae \cite{werner2013, reitberger2012, reitberger2015}, and the still debated WR 11 \cite{pshirkov2016}.\\
In previous studies, the non-thermal photon emission of typical CWB systems was computed using the spectral energy distributions of high-energy particles obtained by combining three-dimensional hydrodynamic numerical simulations with the solution of the transport equation for protons and electrons \cite{reitberger2014a, reitberger2014b}. In these works, particle acceleration is provided by a very simplified term in the transport equation, which is the analytical result obtained when considering a population of suprathermal particles at a shock of infinite extent. The injection efficiency of thermal particles into the acceleration process, i.e. how many of the particles from the thermal distribution are able to recross the shock from downstream to upstream and be injected into DSA, is necessarily prescribed ``by hand''.\\
For studying the microphysics of shock formation and injection efficiency, (full) particle in cell (PIC), and hybrid PIC simulations are suited best: particles move in the simulation box and are scattered by self-consistently generated magnetic turbulences (e.g. \cite{gargate2012, caprioli2014} and references therein). Unfortunately, these approaches are computationally very demanding, and both box sizes and time intervals that can be investigated are limited.\\
Particle acceleration has also been studied by means of Monte Carlo simulations (e.g. \cite{kirk1987,ostrowski1991,baring1993}). Here, particles move undisturbed on a predefined background until a scattering occurs. The scattering process is necessarily modelled, but injection efficiencies can be obtained, depending on the prescribed scattering laws. Furthermore, this simplification significantly reduces the computational load, and allows for much larger regions and time intervals (and in turn energy ranges) to be considered.\\
For this reason, we chose the Monte Carlo method for our purpose of simulating particle acceleration in extended regions, and over wide energy and time intervals. In particular, we employ a test-particle approach, resembling methods used for studying particle injection at oblique shocks (e.g. \cite{baring1993, baring1994, ellison1995}).\\
In order to obtain more realistic background conditions, we use magnetohydrodynamic (MHD) simulations of a typical CWB system.\\ 
In the following, we will first describe the numerical method employed in this work. We will then show the details of the system studied, as well as the results of the simulations. The last section is devoted to the conclusions.

\section{NUMERICAL METHOD}
\begin{figure}[h]
\tabcolsep7pt\begin{tabular}{cc}
\centering
\includegraphics[width=.33\columnwidth]{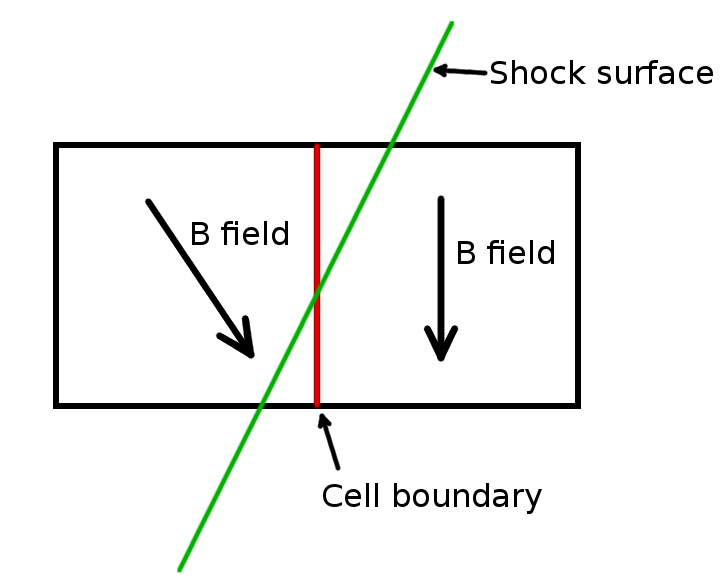}
\caption{Schematic representation of an upstream and a downstream cell, together with the actual shock surface resulting from MHD simulations. On the Cartesian grid, the cell boundary, which divides upstream and downstream, is in general not aligned with the shock surface.}
\label{fig:realshock}
\end{tabular}
\end{figure}

In this work, we investigate particle acceleration in the wind collision region of CWBs. We employ the results of MHD simulations of a typical CWB system, carried out using the \textsc{cronos} code \cite{kissmann2009, kissmann2016}, to characterize the background (plasma flow velocity, magnetic field, electric field, temperature, and density) that influence the particle motion in the simulation box. The background is initialized at the beginning of each simulation and, since we are applying a test-particle approach, it is not changed at runtime by the accelerated particles.\\
In the MHD simulations, the shock front is about three cells wide for numerical reasons only. Moreover, if the shock front is not ``vertical'' in the simulation box, the cell boundary is not parallel to the actual shock front (see Fig.~\ref{fig:realshock}). This can lead to unrealistic results, especially if the gyroradius of the particle is smaller than the cell size, and the particle gyrates many times along the shock surface when encountering the shock. After illustrating the Monte Carlo method itself, we will describe how we deal with these issues.

\subsection{Monte Carlo method}

Protons are injected into the selected cell, with an isotropic Maxwell-Boltzmann distribution of velocities in the frame comoving with the local plasma flow, with the temperature of the local background. Each particle is followed in the frame where the shock front is stationary, and moves on the background electromagnetic field, driven by the Lorentz force, until: (i) a scattering occurs, (ii) it reaches the boundary of a cell, or (iii) it is removed from the system. The time between scatterings is exponentially distributed, with mean value $t_c = \eta \tilde r_g/v$, where $\eta$ is a proportionality factor, $v$ is the speed of the particle, and $\tilde r_g=p/qB$ \cite{ellison1995}. For $\eta=1$ this corresponds to Bohm diffusion. We assume the scattering to be elastic in the local plasma frame. The new $\mu_p=\cos{\theta_p}$, where $\theta_p$ is the pitch angle in the plasma frame, is a random number between 0 and 1, drawn from a uniform distribution. This models large fluctuations in the magnetic field \cite{ellison1995}.\\
When a particle crosses the boundary of a cell, the background changes, and its trajectory is computed accordingly. Particles are removed from the system either after they were scattered $10^6$ times (usually the case for particles which were not accelerated at the shock front and are being advected away by the plasma) or when they reach the outer boundary of the entire simulation box. \\
In order to improve statistics at higher energies, the technique of particle splitting is employed. A statistical weight is assigned to each particle at the beginning of the simulation. When the energy of a particle increases by a factor of 10, its statistical weight is halved, and the particle is splitted into two instantaneously identical ones, which will follow different paths from that moment onwards. The spectrum of the flux through a certain surface is recorded by adding to the appropriate energy bin the statistical weight of the particle crossing it. We verified our code by reproducing the results obtained in Ref. \cite{baring1993}.\\
The background in our simulation is divided into two regions: one includes the shock-front cells, identified as being part of the shock front, the other one comprises all the other cells. When particles are in the non-shock-front cells, the background is just the one resulting from the MHD simulations. When, on the other hand, a particle enters the area marked as being part of the shock front, a different setup is used, with two bigger cells -upstream and downstream-, the background of which is obtained as described in the following section. The position of the particle as it enters this new regime is recorded, so that, when leaving the super-cell regime, the new position in the normal background is found by adding the displacement in the appropriate reference system to the recorded coordinates (see Fig.~\ref{fig:bg} (b)).
\\

 \subsection{Background}

\begin{figure}[h]
\tabcolsep7pt\begin{tabular}{cc}
\centering
\includegraphics[width=.43\columnwidth]{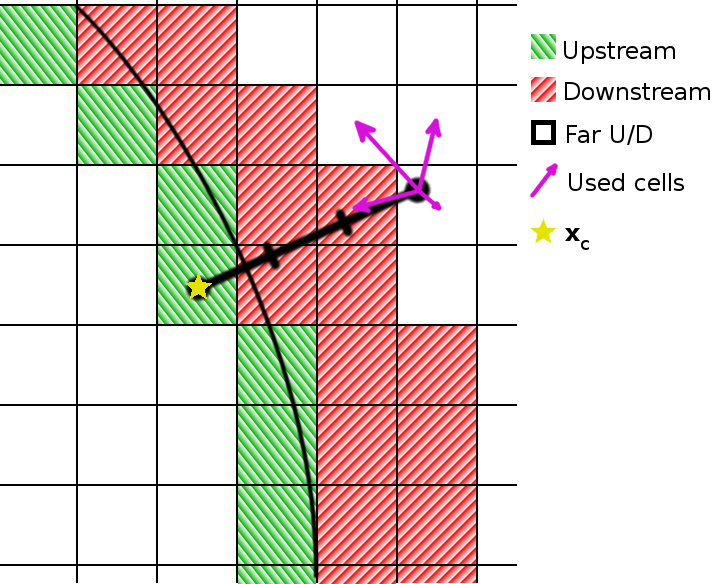}
  & \includegraphics[width=.43\columnwidth]{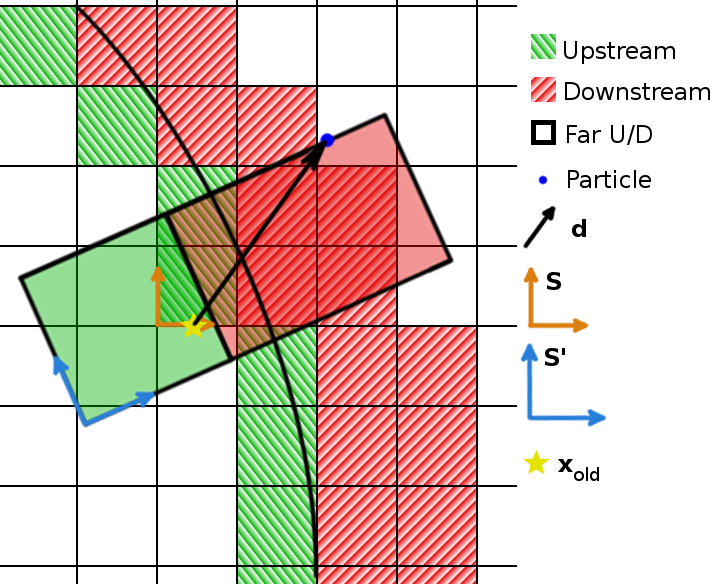}  \\
  (a) \qquad \qquad & (b) \qquad
\caption{(a) Illustration of the method used for the initialization of the background of the downstream super-cells, which results from a weighted average of the fields within a distance $\Delta x$ from the point 3$\Delta x$ downstream of the centre $x_c$ of the upstream shock-front cell, in the direction normal to the shock front ($\Delta x$ is the cell size). (b) Schematic representation of a pair of upstream and downstream super-cells. When a particle enters a cell marked as ``shock front'' cell, its position $x_{old}$ in the simulation domain is recorded, so that, when leaving the super-cell regime, the new position in the normal background is found by adding the displacement vector $\vec{d}$ to the recorded coordinates. $\bf{S}$ is the non-rotated reference frame, $\bf{S'}$ is the rotated reference frame used in the super-cell regime. Cell sizes are exaggerated for display purposes; super-cell sizes are $(2\Delta x)^3$ upstream, and $3\Delta x \times 2\Delta x \times 2\Delta x $ downstream.}
\label{fig:bg}
\end{tabular}
\end{figure}

The background is obtained from MHD simulations. At the shock fronts, the shock transition layer is about 3 cells wide, and the cell surface is in general not aligned with the shock front. We therefore divided the computational domain into two regions: ``shock front'' and ``normal'' cells. In the shock front domain, a system of ``superimposed'' cells (super-cells) is used. For technical reasons, these have dimensions $(2\Delta x)^3$ upstream, and $3\Delta x \times (2\Delta x)^2$ downstream. In order to initialize the super-cells, one first needs to locate the position of the shock front. This is done by setting a threshold for the temperature: at the shock fronts the plasma temperature abruptly rises from $\sim$ 10$^4$K to $\sim 10^7$-$10^8$K. The temperature gradient was found to be a good tracer of the shock front position by \citet{reitberger2014a}.\\
Every upstream super-cell is associated with a downstream one (and vice versa). The background of the upstream super-cells is obtained by averaging the fields of the neighbouring upstream shock-front cells. Following \citet{reitberger2014a}, the background of the associated downstream super-cell is obtained by computing a weighted average of the fields within a distance \mbox{$\Delta x$ ($ \equiv$ cell size)} from the point 3$\Delta x$ downstream of the upstream shock-front cell, in the direction normal to the shock front (see Fig.~\ref{fig:bg} (a)). The vector fields (flow velocity, magnetic field, electric field) are then rotated, so that shock normal and normal to the surface between upstream and downstream super-cells are parallel (reference frame $\bf{S'}$ in Fig.~\ref{fig:bg} (b)).

\section{RESULTS}

\subsection{Parameters of the simulations}
 In the following we consider a B star and Wolf-Rayet star binary system, with the parameters listed in Table~\ref{tab:stellarwind}. 

\begin{table}[h!]
\caption{Stellar and wind parameters of a typical colliding-wind binary system, as in \citet{kissmann2016}. $M_{*}$ is the stellar mass, $R_{*}$ the stellar radius, $T_{*}$	the effective temperature, $L_{*}$ the luminosity, $\dot{M}$ the mass loss rate, $v_{\infty}$ the terminal velocity of the wind, and $B_{*}$ the surface magnetic field.
}
		\label{tab:stellarwind}
		\centering
		\begin{tabular}{l*{7}{c}}

		\hline
		Star 	& $M_{*}$ 				& $R_{*}$				& $T_{*}$	& $L_{*}$				& $\dot{M}$ 					& $v_{\infty}$ 		& $B_{*}$\\
				& [M$_{\astrosun}$]		& [R$_{\astrosun}$] 	& [K]		& [L$_{\astrosun}$]		& [M$_{\astrosun}$ yr$^{-1}$]	& [km s$^{-1}$] 	& [G]\\
		\hline
		B		& $30$					& $20$					& $23000$	& $10^5$				& $10^{-6}$						& 4000				& 100\\
		WR		& $30$					& $10$					& $40000$	& $2.3\times 10^5$		& $10^{-5}$						& 4000				& 100\\
		\hline
		\end{tabular}
		\end{table}

\noindent The stellar separation is $R=1440 \mbox{ R}_{\astrosun}$. The region used in the Monte Carlo simulations consists of ($151\times81\times151$) cubic cells of dimension (3.9$\mbox{ R}_{\astrosun}$)$^3$.
We stress that no analytical prescriptions are necessary concerning the large-scale magnetic field at the WCR, since the magnetic field was evolved dynamically in the MHD background simulations.\\
Particles are injected upstream of the shock fronts, at different positions along the WCR, on the $x$-$z$ plane, at $z=40\ R_{\astrosun}$, $z=-420\ R_{\astrosun}$, and $z=420\ R_{\astrosun}$. The fluxes are recorded when particles cross the shock fronts. Here, we show the results of simulations carried out with $\eta=1$ (highly turbulent medium) \cite{ellison1995}. The impact of varying $\eta$ on the results is subject to future studies.

\subsection{Spectral indices and injection efficiencies}
The analytical result for the density of particles accelerated via DSA, at a shock with compression ratio $r$, yields the well known dependence $n(p)\propto p^{-\sigma}$, where $p$ is the momentum, $n$ is the differential particle density, and \mbox{$\sigma=(r+2)/(r-1)$} is the spectral index. 
The differential current of protons in terms of the kinetic energy $E$ is:
\begin{equation}
J(E) = v n(E) \propto  \left[E\left(E+2 m_p c^2\right)\right]^{-\frac{\sigma}{2}} \ ,
\label{eq:currentEk}
\end{equation}
where $v$ is the speed of the proton, $m_p$ is its rest mass, and $c$ is the speed of light.\\

\begin{figure}[h]
\begin{tabular}{cc}
	\includegraphics[width=0.432\columnwidth]{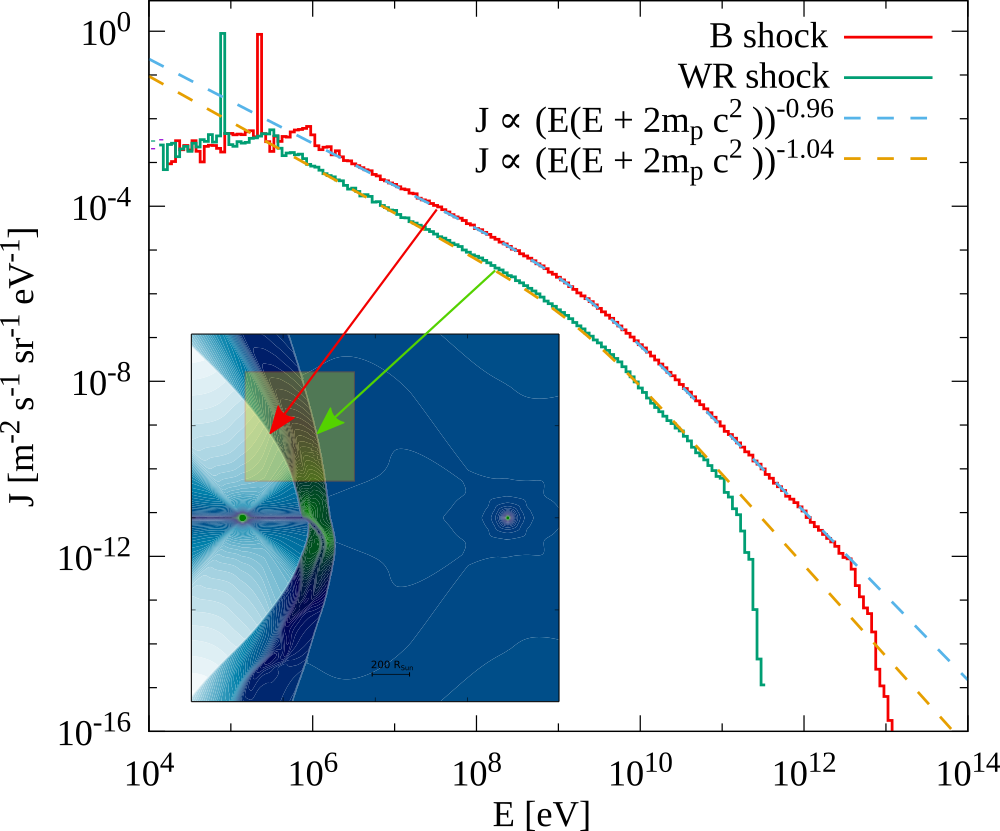}\\
	\qquad (c)
\end{tabular}
\caption{Spectra of fluxes through the shock fronts for protons injected at (a) $z=40\ R_{\astrosun}$, (b) $z=-420\ R_{\astrosun}$, and (c) $z=420\ R_{\astrosun}$. The dashed curves are obtained by fitting the function of Equation \ref{eq:currentEk} to the data. In the blue box we show the plasma's background speed in a cut through the numerical domain at $y=0$. The B star is on the left, the WR star on the right. The bow-shaped region, closer to the B star and bent around it, is the WCR. The smaller boxes represent the sections used for the Monte Carlo simulations.}
\label{fig:spectra}
\end{figure}

\noindent The spectra resulting from the simulations can be seen in Fig.~\ref{fig:spectra}. The magnetic field on the WR-side of the WCR is weaker, which causes a difference of up to two orders of magnitude in the maximal energy reached by particles injected on the WR-side ($E^{WR}_{max}\sim 10^{11}$ eV) and those injected on the B-side ($E^{B}_{max}\sim 10^{12}-10^{13}$ eV). Moreover, the compression ratio is in general higher on the B-side, which results in harder spectra.\\
The Monte Carlo simulations also allow to estimate injection efficiencies for the system, once the scattering law (i.e. scattering operator and mean free path dependence on different parameters) has been chosen, and under the assumption that the shock front is seen as a sharp transition by the protons. If we define the injection efficiency $\varepsilon$ as:
\begin{equation}
\varepsilon = \frac{n_{NT}}{n_{TOT}} \ ,
\label{eq:defeff}
\end{equation}
where $n_{NT}$ and $n_{TOT}$ are the particle density in the non-thermal tail and in the total particle distribution, respectively, we obtain $8\%\leq\varepsilon_{_{WR}}\leq16\%$ for the WR shock, and $9\%\leq\varepsilon_{_{B}}\leq26\%$ for the B shock. The variation in the injection efficiency seems to depend not only on the side of the WCR where the particles are injected, but also on the distance of the injection position from its apex, as can be seen in Table~\ref{tab:results}. This is probably due to both different plasma flow velocities and shock obliquities. The spectra obtained when injecting particles at $z= 420\ R_{\astrosun}$ are harder than those of particles injected at $z=-420\ R_{\astrosun}$. This is ascribable to different compression ratios at the two injection positions. Although an association of this feature with the asymmetry of the plasma flow is tempting, we note that the compression ratio changes not continuously along the shock front, and further simulations are needed in order to verify if spectra are systematically harder above the plane at $z=0$. We note that efficiencies are rather high, therefore including feedback of the accelerated particles on the plasma would yield more realistic results. Nevertheless, our simulations indicate that the change in injection efficiency along the WCR may indeed be relevant when modelling non-thermal emission from CWB systems.




\begin{figure}[h]
\tabcolsep7pt\begin{tabular}{cc}
	\includegraphics[width=0.432\columnwidth]{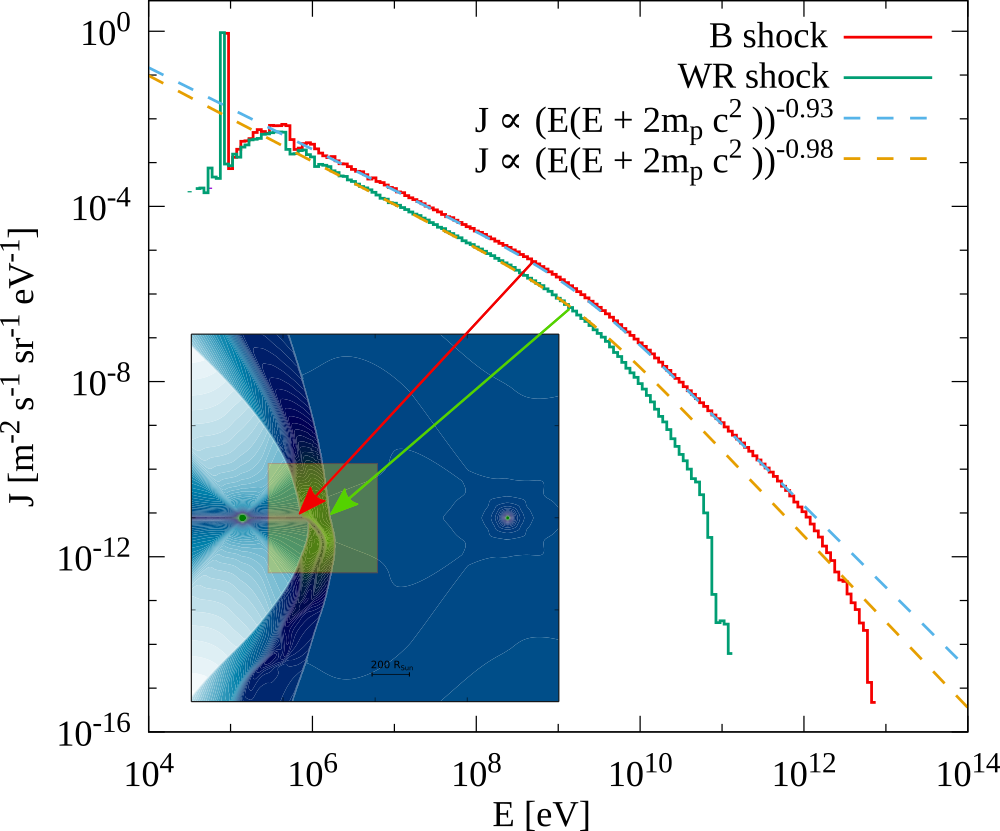}
	&
	\includegraphics[width=0.432\columnwidth]{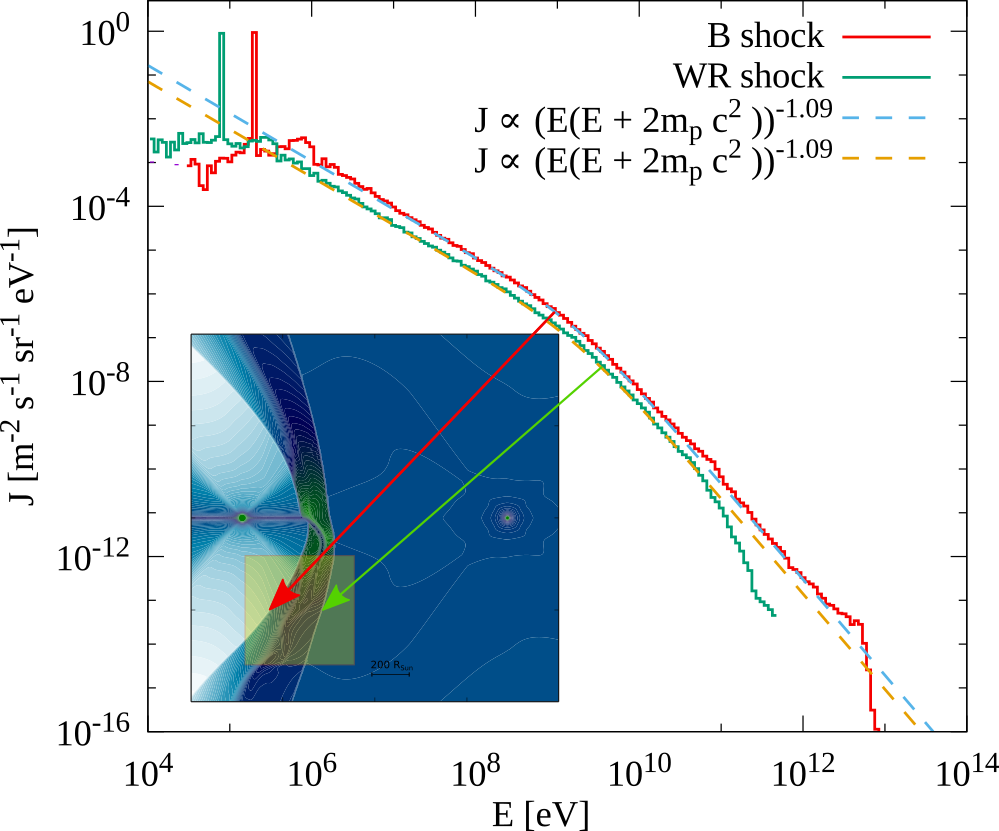}\\
	\qquad (a) & \qquad (b)
\end{tabular}
\end{figure}

\begin{table}[h!]
\caption{Spectral indices and injection efficiencies of protons injected at $z=40\ R_{\astrosun}$, $z=-420\ R_{\astrosun}$, and $z=420\ R_{\astrosun}$. The errors refer to the fit of Equation \ref{eq:currentEk} to the data.}
\label{tab:results}
\centering
\begin{tabular}{*{4}{c}}

\hline
z & Side of WCR & Spectral index	& Injection efficiency \\
$[R_{\astrosun}]$	&	& $\sigma$ 				& $\varepsilon$ \\
\hline
40	&B		& $1.86 \pm 0.02$			& $\approx26\%$\\
	&WR		& $1.96 \pm 0.02$			& $\approx16\%$\\
\\
-420	&B		& $2.19 \pm 0.01$			& $\approx9\%$\\
	&WR		& $2.18 \pm 0.04$			& $\approx8\%$\\
\\
420	&B		& $1.93 \pm 0.02$			& $\approx16\%$\\
	&WR		& $2.08 \pm 0.04$			& $\approx13\%$\\
\hline
\end{tabular}
\end{table}

\pagebreak
\section{CONCLUSIONS}

In this work, we investigated the acceleration of protons in a typical colliding-wind binary system. Our Monte Carlo test-particle simulations employ the results of magnetohydrodynamic simulations, which determine the background on which protons move.\\
We found a difference in both the spectral index and the highest energy reached by the accelerated particles, depending on the considered side of the WCR, due to different compression ratios and strengths of the magnetic field. Moreover, we found that these two characteristics of the spectra of non-thermal protons also vary moving away from the apex of the WCR.
A similar remark can be done concerning the injection efficiencies, which seem to be in general higher on the B-side of the WCR, and to decrease further away from its apex, along the shock fronts. Our results indicate that a variation of the injection efficiency probably needs to be taken into account in models aiming at predicting $\gamma$-ray fluxes produced by non-thermal particles accelerated in CWB systems.


\section{ACKNOWLEDGMENTS}
E. G. and A. R. acknowledge financial support from the Austrian Science Fund (FWF), project \mbox{P 24926-N27.}


\nocite{*}
\bibliographystyle{aipnum-cp}%
\bibliography{cmhdmc}%

\begin{thebibliography}{23}%
\makeatletter
\providecommand \@ifxundefined [1]{%
 \@ifx{#1\undefined}
}%
\providecommand \@ifnum [1]{%
 \ifnum #1\expandafter \@firstoftwo
 \else \expandafter \@secondoftwo
 \fi
}%
\providecommand \@ifx [1]{%
 \ifx #1\expandafter \@firstoftwo
 \else \expandafter \@secondoftwo
 \fi
}%
\providecommand \natexlab [1]{#1}%
\providecommand \enquote  [1]{``#1''}%
\providecommand \bibnamefont  [1]{#1}%
\providecommand \bibfnamefont [1]{#1}%
\providecommand \citenamefont [1]{#1}%
\providecommand \href@noop [0]{\@secondoftwo}%
\providecommand \href [0]{\begingroup \@sanitize@url \@href}%
\providecommand \@href[1]{\@@startlink{#1}\@@href}%
\providecommand \@@href[1]{\endgroup#1\@@endlink}%
\providecommand \@sanitize@url [0]{\catcode `\$12\catcode `\&12\catcode
  `\#12\catcode `\^12\catcode `\_12\catcode `\%12\relax}%
\providecommand \@@startlink[1]{}%
\providecommand \@@endlink[0]{}%
\providecommand \url  [0]{\begingroup\@sanitize@url \@url }%
\providecommand \@url [1]{\endgroup\@href {#1}{\urlprefix }}%
\providecommand \urlprefix  [0]{URL }%
\providecommand \Eprint [0]{\href }%
\providecommand \doibase [0]{http://dx.doi.org/}%
\providecommand \selectlanguage [0]{\@gobble}%
\providecommand \bibinfo  [0]{\@secondoftwo}%
\providecommand \bibfield  [0]{\@secondoftwo}%
\providecommand \translation [1]{[#1]}%
\providecommand \BibitemOpen [0]{}%
\providecommand \bibitemStop [0]{}%
\providecommand \bibitemNoStop [0]{.\EOS\space}%
\providecommand \EOS [0]{\spacefactor3000\relax}%
\providecommand \BibitemShut  [1]{\csname bibitem#1\endcsname}%
\let\auto@bib@innerbib\@empty
\bibitem [{\citenamefont {Jones}\ and\ \citenamefont
  {Ellison}(1991)}]{jones1991rev}%
  \BibitemOpen
  \bibfield  {author} {\bibinfo {author} {\bibfnamefont {F.~C.}\ \bibnamefont
  {Jones}}\ and\ \bibinfo {author} {\bibfnamefont {D.~C.}\ \bibnamefont
  {Ellison}},\ }\href@noop {} {\bibfield  {journal} {\bibinfo  {journal} {Space
  Sci. Rev.}\ }\textbf {\bibinfo {volume} {58}},\ \unskip\ \bibinfo {pages}
  {259--346} (\bibinfo {year} {1991})}\BibitemShut {NoStop}%
\bibitem [{\citenamefont {Drury}(1983)}]{drury1983}%
  \BibitemOpen
  \bibfield  {author} {\bibinfo {author} {\bibfnamefont {L.~O.}\ \bibnamefont
  {Drury}},\ }\href@noop {} {\bibfield  {journal} {\bibinfo  {journal} {Rep.
  Prog. Phys.}\ }\textbf {\bibinfo {volume} {46}},\ p.\ \bibinfo {pages} {973}
  (\bibinfo {year} {1983})}\BibitemShut {NoStop}%
\bibitem [{\citenamefont {Williams}\ \emph {et~al.}(1997)\citenamefont
  {Williams}, \citenamefont {Dougherty}, \citenamefont {Davis}, \citenamefont
  {van~der Hucht}, \citenamefont {Bode},\ and\ \citenamefont
  {Gunawan}}]{williams1997}%
  \BibitemOpen
  \bibfield  {author} {\bibinfo {author} {\bibfnamefont {P.~M.}\ \bibnamefont
  {Williams}}, \bibinfo {author} {\bibfnamefont {S.~M.}\ \bibnamefont
  {Dougherty}}, \bibinfo {author} {\bibfnamefont {R.~J.}\ \bibnamefont
  {Davis}}, \bibinfo {author} {\bibfnamefont {K.~A.}\ \bibnamefont {van~der
  Hucht}}, \bibinfo {author} {\bibfnamefont {M.~F.}\ \bibnamefont {Bode}}, \
  and\ \bibinfo {author} {\bibfnamefont {D.~Y. A.~S.}\ \bibnamefont
  {Gunawan}},\ }\href@noop {} {\bibfield  {journal} {\bibinfo  {journal} {Mon.
  Not. R. Astron. Soc.}\ }\textbf {\bibinfo {volume} {289}},\ \unskip\ \bibinfo
  {pages} {10--20} (\bibinfo {year} {1997})}\BibitemShut {NoStop}%
\bibitem [{\citenamefont {Chapman}\ \emph {et~al.}(1999)\citenamefont
  {Chapman}, \citenamefont {Leitherer}, \citenamefont {Koribalski},
  \citenamefont {Bouter},\ and\ \citenamefont {Storey}}]{chapman1999}%
  \BibitemOpen
  \bibfield  {author} {\bibinfo {author} {\bibfnamefont {J.~M.}\ \bibnamefont
  {Chapman}}, \bibinfo {author} {\bibfnamefont {C.}~\bibnamefont {Leitherer}},
  \bibinfo {author} {\bibfnamefont {B.}~\bibnamefont {Koribalski}}, \bibinfo
  {author} {\bibfnamefont {R.}~\bibnamefont {Bouter}}, \ and\ \bibinfo {author}
  {\bibfnamefont {M.}~\bibnamefont {Storey}},\ }\href@noop {} {\bibfield
  {journal} {\bibinfo  {journal} {Astrophys. J.}\ }\textbf {\bibinfo {volume}
  {518}},\ p.\ \bibinfo {pages} {890} (\bibinfo {year} {1999})}\BibitemShut
  {NoStop}%
\bibitem [{\citenamefont {Dougherty}\ \emph {et~al.}(2005)\citenamefont
  {Dougherty}, \citenamefont {Beasley}, \citenamefont {Claussen}, \citenamefont
  {Zauderer},\ and\ \citenamefont {Bolingbroke}}]{dougherty2005}%
  \BibitemOpen
  \bibfield  {author} {\bibinfo {author} {\bibfnamefont {S.~M.}\ \bibnamefont
  {Dougherty}}, \bibinfo {author} {\bibfnamefont {A.~J.}\ \bibnamefont
  {Beasley}}, \bibinfo {author} {\bibfnamefont {M.~J.}\ \bibnamefont
  {Claussen}}, \bibinfo {author} {\bibfnamefont {B.~A.}\ \bibnamefont
  {Zauderer}}, \ and\ \bibinfo {author} {\bibfnamefont {N.~J.}\ \bibnamefont
  {Bolingbroke}},\ }\href@noop {} {\bibfield  {journal} {\bibinfo  {journal}
  {Astrophys. J.}\ }\textbf {\bibinfo {volume} {623}},\ p.\ \bibinfo {pages}
  {447} (\bibinfo {year} {2005})}\BibitemShut {NoStop}%
\bibitem [{\citenamefont {Eichler}\ and\ \citenamefont
  {Usov}(1993)}]{eichler1993}%
  \BibitemOpen
  \bibfield  {author} {\bibinfo {author} {\bibfnamefont {D.}~\bibnamefont
  {Eichler}}\ and\ \bibinfo {author} {\bibfnamefont {V.~V.}\ \bibnamefont
  {Usov}},\ }\href@noop {} {\bibfield  {journal} {\bibinfo  {journal}
  {Astrophys. J.}\ }\textbf {\bibinfo {volume} {402}},\ \unskip\ \bibinfo
  {pages} {271--279} (\bibinfo {year} {1993})}\BibitemShut {NoStop}%
\bibitem [{\citenamefont {Reimer}, \citenamefont {Pohl},\ and\ \citenamefont
  {Reimer}(2006)}]{reimer2006}%
  \BibitemOpen
  \bibfield  {author} {\bibinfo {author} {\bibfnamefont {A.}~\bibnamefont
  {Reimer}}, \bibinfo {author} {\bibfnamefont {M.}~\bibnamefont {Pohl}}, \ and\
  \bibinfo {author} {\bibfnamefont {O.}~\bibnamefont {Reimer}},\ }\href@noop {}
  {\bibfield  {journal} {\bibinfo  {journal} {Astrophys. J.}\ }\textbf
  {\bibinfo {volume} {644}},\ p.\ \bibinfo {pages} {1118} (\bibinfo {year}
  {2006})}\BibitemShut {NoStop}%
\bibitem [{\citenamefont {Benaglia}\ and\ \citenamefont
  {Romero}(2003)}]{benaglia2003}%
  \BibitemOpen
  \bibfield  {author} {\bibinfo {author} {\bibfnamefont {P.}~\bibnamefont
  {Benaglia}}\ and\ \bibinfo {author} {\bibfnamefont {G.~E.}\ \bibnamefont
  {Romero}},\ }\href@noop {} {\bibfield  {journal} {\bibinfo  {journal}
  {Astron. Astrophys.}\ }\textbf {\bibinfo {volume} {399}},\ \unskip\ \bibinfo
  {pages} {1121--1134} (\bibinfo {year} {2003})}\BibitemShut {NoStop}%
\bibitem [{\citenamefont {Werner}\ \emph {et~al.}(2013)\citenamefont {Werner},
  \citenamefont {Reimer}, \citenamefont {Reimer},\ and\ \citenamefont
  {Egberts}}]{werner2013}%
  \BibitemOpen
  \bibfield  {author} {\bibinfo {author} {\bibfnamefont {M.}~\bibnamefont
  {Werner}}, \bibinfo {author} {\bibfnamefont {O.}~\bibnamefont {Reimer}},
  \bibinfo {author} {\bibfnamefont {A.}~\bibnamefont {Reimer}}, \ and\ \bibinfo
  {author} {\bibfnamefont {K.}~\bibnamefont {Egberts}},\ }\href@noop {}
  {\bibfield  {journal} {\bibinfo  {journal} {Astron. Astrophys.}\ }\textbf
  {\bibinfo {volume} {555}},\ p.\ \bibinfo {pages} {A102} (\bibinfo {year}
  {2013})}\BibitemShut {NoStop}%
\bibitem [{\citenamefont {Reitberger}\ \emph {et~al.}(2012)\citenamefont
  {Reitberger}, \citenamefont {Reimer}, \citenamefont {Reimer}, \citenamefont
  {Werner}, \citenamefont {Egberts},\ and\ \citenamefont
  {Takahashi}}]{reitberger2012}%
  \BibitemOpen
  \bibfield  {author} {\bibinfo {author} {\bibfnamefont {K.}~\bibnamefont
  {Reitberger}}, \bibinfo {author} {\bibfnamefont {O.}~\bibnamefont {Reimer}},
  \bibinfo {author} {\bibfnamefont {A.}~\bibnamefont {Reimer}}, \bibinfo
  {author} {\bibfnamefont {M.}~\bibnamefont {Werner}}, \bibinfo {author}
  {\bibfnamefont {K.}~\bibnamefont {Egberts}}, \ and\ \bibinfo {author}
  {\bibfnamefont {H.}~\bibnamefont {Takahashi}},\ }\href@noop {} {\bibfield
  {journal} {\bibinfo  {journal} {Astron. Astrophys.}\ }\textbf {\bibinfo
  {volume} {544}},\ p.\ \bibinfo {pages} {A98} (\bibinfo {year}
  {2012})}\BibitemShut {NoStop}%
\bibitem [{\citenamefont {{Reitberger}}\ \emph {et~al.}(2015)\citenamefont
  {{Reitberger}}, \citenamefont {{Reimer}}, \citenamefont {{Reimer}},\ and\
  \citenamefont {{Takahashi}}}]{reitberger2015}%
  \BibitemOpen
  \bibfield  {author} {\bibinfo {author} {\bibfnamefont {K.}~\bibnamefont
  {{Reitberger}}}, \bibinfo {author} {\bibfnamefont {A.}~\bibnamefont
  {{Reimer}}}, \bibinfo {author} {\bibfnamefont {O.}~\bibnamefont {{Reimer}}},
  \ and\ \bibinfo {author} {\bibfnamefont {H.}~\bibnamefont {{Takahashi}}},\
  }\href@noop {} {\bibfield  {journal} {\bibinfo  {journal} {Astron.
  Astrophys.}\ }\textbf {\bibinfo {volume} {577}},\ p.\ \bibinfo {pages} {A100}
  (\bibinfo {year} {2015})}\BibitemShut {NoStop}%
\bibitem [{\citenamefont {{Pshirkov}}(2016)}]{pshirkov2016}%
  \BibitemOpen
  \bibfield  {author} {\bibinfo {author} {\bibfnamefont {M.~S.}\ \bibnamefont
  {{Pshirkov}}},\ }\href@noop {} {\bibfield  {journal} {\bibinfo  {journal}
  {Mon. Not. R. Astron. Soc.}\ }\textbf {\bibinfo {volume} {457}},\ \unskip\
  \bibinfo {pages} {L99--L102} (\bibinfo {year} {2016})}\BibitemShut {NoStop}%
\bibitem [{\citenamefont {Reitberger}\ \emph {et~al.}(014a)\citenamefont
  {Reitberger}, \citenamefont {Kissmann}, \citenamefont {Reimer}, \citenamefont
  {Reimer},\ and\ \citenamefont {Dubus}}]{reitberger2014a}%
  \BibitemOpen
  \bibfield  {author} {\bibinfo {author} {\bibfnamefont {K.}~\bibnamefont
  {Reitberger}}, \bibinfo {author} {\bibfnamefont {R.}~\bibnamefont
  {Kissmann}}, \bibinfo {author} {\bibfnamefont {A.}~\bibnamefont {Reimer}},
  \bibinfo {author} {\bibfnamefont {O.}~\bibnamefont {Reimer}}, \ and\ \bibinfo
  {author} {\bibfnamefont {G.}~\bibnamefont {Dubus}},\ }\href@noop {}
  {\bibfield  {journal} {\bibinfo  {journal} {Astrophys. J.}\ }\textbf
  {\bibinfo {volume} {782}},\ p.~\bibinfo {pages} {96} (\bibinfo {year}
  {2014a})}\BibitemShut {NoStop}%
\bibitem [{\citenamefont {Reitberger}\ \emph {et~al.}(014b)\citenamefont
  {Reitberger}, \citenamefont {Kissmann}, \citenamefont {Reimer},\ and\
  \citenamefont {Reimer}}]{reitberger2014b}%
  \BibitemOpen
  \bibfield  {author} {\bibinfo {author} {\bibfnamefont {K.}~\bibnamefont
  {Reitberger}}, \bibinfo {author} {\bibfnamefont {R.}~\bibnamefont
  {Kissmann}}, \bibinfo {author} {\bibfnamefont {A.}~\bibnamefont {Reimer}}, \
  and\ \bibinfo {author} {\bibfnamefont {O.}~\bibnamefont {Reimer}},\
  }\href@noop {} {\bibfield  {journal} {\bibinfo  {journal} {Astrophys. J.}\
  }\textbf {\bibinfo {volume} {789}},\ p.~\bibinfo {pages} {87} (\bibinfo
  {year} {2014b})}\BibitemShut {NoStop}%
\bibitem [{\citenamefont {{Gargat{\'e}}}\ and\ \citenamefont
  {{Spitkovsky}}(2012)}]{gargate2012}%
  \BibitemOpen
  \bibfield  {author} {\bibinfo {author} {\bibfnamefont {L.}~\bibnamefont
  {{Gargat{\'e}}}}\ and\ \bibinfo {author} {\bibfnamefont {A.}~\bibnamefont
  {{Spitkovsky}}},\ }\href@noop {} {\bibfield  {journal} {\bibinfo  {journal}
  {Astrophys. J.}\ }\textbf {\bibinfo {volume} {744}},\ p.~\bibinfo {pages}
  {67} (\bibinfo {year} {2012})}\BibitemShut {NoStop}%
\bibitem [{\citenamefont {{Caprioli}}\ and\ \citenamefont
  {{Spitkovsky}}(2014)}]{caprioli2014}%
  \BibitemOpen
  \bibfield  {author} {\bibinfo {author} {\bibfnamefont {D.}~\bibnamefont
  {{Caprioli}}}\ and\ \bibinfo {author} {\bibfnamefont {A.}~\bibnamefont
  {{Spitkovsky}}},\ }\href@noop {} {\bibfield  {journal} {\bibinfo  {journal}
  {Astrophys. J.}\ }\textbf {\bibinfo {volume} {783}},\ p.~\bibinfo {pages}
  {91} (\bibinfo {year} {2014})}\BibitemShut {NoStop}%
\bibitem [{\citenamefont {{Kirk}}\ and\ \citenamefont
  {{Schneider}}(1987)}]{kirk1987}%
  \BibitemOpen
  \bibfield  {author} {\bibinfo {author} {\bibfnamefont {J.~G.}\ \bibnamefont
  {{Kirk}}}\ and\ \bibinfo {author} {\bibfnamefont {P.}~\bibnamefont
  {{Schneider}}},\ }\href@noop {} {\bibfield  {journal} {\bibinfo  {journal}
  {Astrophys. J.}\ }\textbf {\bibinfo {volume} {322}},\ \unskip\ \bibinfo
  {pages} {256--265} (\bibinfo {year} {1987})}\BibitemShut {NoStop}%
\bibitem [{\citenamefont {{Ostrowski}}(1991)}]{ostrowski1991}%
  \BibitemOpen
  \bibfield  {author} {\bibinfo {author} {\bibfnamefont {M.}~\bibnamefont
  {{Ostrowski}}},\ }\href@noop {} {\bibfield  {journal} {\bibinfo  {journal}
  {Mon. Not. R. Astron. Soc.}\ }\textbf {\bibinfo {volume} {249}},\ \unskip\
  \bibinfo {pages} {551--559} (\bibinfo {year} {1991})}\BibitemShut {NoStop}%
\bibitem [{\citenamefont {Baring}, \citenamefont {Ellison},\ and\ \citenamefont
  {Jones}(1993)}]{baring1993}%
  \BibitemOpen
  \bibfield  {author} {\bibinfo {author} {\bibfnamefont {M.~G.}\ \bibnamefont
  {Baring}}, \bibinfo {author} {\bibfnamefont {D.~C.}\ \bibnamefont {Ellison}},
  \ and\ \bibinfo {author} {\bibfnamefont {F.~C.}\ \bibnamefont {Jones}},\
  }\href@noop {} {\bibfield  {journal} {\bibinfo  {journal} {Astrophys. J.}\
  }\textbf {\bibinfo {volume} {409}},\ \unskip\ \bibinfo {pages} {327--332}
  (\bibinfo {year} {1993})}\BibitemShut {NoStop}%
\bibitem [{\citenamefont {{Baring}}, \citenamefont {{Ellison}},\ and\
  \citenamefont {{Jones}}(1994)}]{baring1994}%
  \BibitemOpen
  \bibfield  {author} {\bibinfo {author} {\bibfnamefont {M.~G.}\ \bibnamefont
  {{Baring}}}, \bibinfo {author} {\bibfnamefont {D.~C.}\ \bibnamefont
  {{Ellison}}}, \ and\ \bibinfo {author} {\bibfnamefont {F.~C.}\ \bibnamefont
  {{Jones}}},\ }\href@noop {} {\bibfield  {journal} {\bibinfo  {journal}
  {Astrophys. J.}\ }\textbf {\bibinfo {volume} {90}},\ \unskip\ \bibinfo
  {pages} {547--552} (\bibinfo {year} {1994})}\BibitemShut {NoStop}%
\bibitem [{\citenamefont {{Ellison}}, \citenamefont {{Baring}},\ and\
  \citenamefont {{Jones}}(1995)}]{ellison1995}%
  \BibitemOpen
  \bibfield  {author} {\bibinfo {author} {\bibfnamefont {D.~C.}\ \bibnamefont
  {{Ellison}}}, \bibinfo {author} {\bibfnamefont {M.~G.}\ \bibnamefont
  {{Baring}}}, \ and\ \bibinfo {author} {\bibfnamefont {F.~C.}\ \bibnamefont
  {{Jones}}},\ }\href@noop {} {\bibfield  {journal} {\bibinfo  {journal}
  {Astrophys. J.}\ }\textbf {\bibinfo {volume} {453}},\ p.\ \bibinfo {pages}
  {873} (\bibinfo {year} {1995})}\BibitemShut {NoStop}%
\bibitem [{\citenamefont {Kissmann}, \citenamefont {Pomoell},\ and\
  \citenamefont {Kley}(2009)}]{kissmann2009}%
  \BibitemOpen
  \bibfield  {author} {\bibinfo {author} {\bibfnamefont {R.}~\bibnamefont
  {Kissmann}}, \bibinfo {author} {\bibfnamefont {J.}~\bibnamefont {Pomoell}}, \
  and\ \bibinfo {author} {\bibfnamefont {W.}~\bibnamefont {Kley}},\ }\href@noop
  {} {\bibfield  {journal} {\bibinfo  {journal} {J. Comput. Phys.}\ }\textbf
  {\bibinfo {volume} {228}},\ \unskip\ \bibinfo {pages} {2119--2131} (\bibinfo
  {year} {2009})}\BibitemShut {NoStop}%
\bibitem [{\citenamefont {{Kissmann}}\ \emph {et~al.}(2016)\citenamefont
  {{Kissmann}}, \citenamefont {{Reitberger}}, \citenamefont {{Reimer}},
  \citenamefont {{Reimer}},\ and\ \citenamefont {{Grimaldo}}}]{kissmann2016}%
  \BibitemOpen
  \bibfield  {author} {\bibinfo {author} {\bibfnamefont {R.}~\bibnamefont
  {{Kissmann}}}, \bibinfo {author} {\bibfnamefont {K.}~\bibnamefont
  {{Reitberger}}}, \bibinfo {author} {\bibfnamefont {O.}~\bibnamefont
  {{Reimer}}}, \bibinfo {author} {\bibfnamefont {A.}~\bibnamefont {{Reimer}}},
  \ and\ \bibinfo {author} {\bibfnamefont {E.}~\bibnamefont {{Grimaldo}}},\
  }\href@noop {} {\bibfield  {journal} {\bibinfo  {journal} {Accepted for
  publication Astrophys. J.}\ } (\bibinfo {year} {2016})},\ \Eprint
  {http://arxiv.org/abs/1609.01130} {arXiv:1609.01130} \BibitemShut {NoStop}%
\end{thebibliography}%

\end{document}